\documentstyle[aps,epsfig,multicol]{revtex}

\begin{document}

\newcommand{\onecolm}{
   \end{multicols}
   \vspace{-1.5\baselineskip}
   \vspace{-\parskip}
   \noindent\rule{0.5\textwidth}{0.1ex}\rule{0.1ex}{2ex}\hfill 
	}
\newcommand{\twocolm}{
   \hfill\raisebox{-1.9ex}{\rule{0.1ex}{2ex}}\rule{0.5\textwidth}{0.1ex}
   \vspace{-1.5\baselineskip}
   \vspace{-\parskip}
   \begin{multicols}{2}
	}

\renewcommand{\theequation}{\arabic{equation}}
\def\bra#1{\left\langle{#1}\,\right|\,}    
\def\ket#1{\,\left|\,{#1}\right\rangle}    

\title{Positive cross-correlations induced by ferromagnetic 
contacts}
\author{Fabio Taddei$^{(1,2)}$\thanks{e-mail: Fabio.Taddei@dmfci.unict.it} 
and
Rosario Fazio$^{(3)}$}
\address{$^1$ISI Foundation, Viale Settimio Severo, 65, I-10133  Torino, Italy\\
$^2$NEST-INFM \& Dipartimento di Metodologie Fisiche e Chimiche (DMFCI), 
Universit\`a di Catania, Viale A. Doria, 6, I-95125 Catania, Italy\\
$^3$NEST-INFM \& Scuola Normale Superiore, I-56126 Pisa, Italy}
\date{\today}
\maketitle

\begin{abstract}
Due to the 
Fermionic nature of carriers,
correlations between electric currents flowing through two different 
contacts attached to a conductor present a negative sign.
Possibility for positive cross-correlations has been demonstrated in 
hybrid normal/superconductor structures under certain conditions.
In this paper we show that positive cross-correlations can be
induced, if not already present, in such structures by employing
ferromagnetic leads with 
magnetizations aligned anti-parallel to each other.
We consider three-terminal hybrid structures and calculate the mean-square
correlations of current fluctuations as a function of the bias voltage at
finite temperature.
\end{abstract}

\begin{multicols}{2}

\section{Introduction}
Due to the discrete nature of charge carriers, the
electronic current flowing through a conductor is subject to
time-dependent fluctuations around its mean value.
Correlations of such fluctuations are of great interest since they provide
more information, with respect to average currents, on the physics of transport.
Shot noise, defined as the mean-square
fluctuations of the current flowing through a given
terminal at zero temperature, has been extensively studied in a wide variety
of systems (for a review on the subject see Ref. \cite{But.Bl}).
Here we are interested in the correlations of
the current fluctuations between different contacts
(cross-correlations).
B\"uttiker \cite{But.1} first pointed out that while cross-correlations
can either be positive or negative for Bosons, for Fermions they are
necessarily negative, both in the equilibrium situation and in the transport
regime (see also Ref. \cite{Ma.La}).

At this point a question arises naturally: what happens in the presence of
superconductivity, where charge current is carried by the superconducting
condensate instead of by Fermionic excitations.
In the sub-gap regime the presence of the condensate manifests itself in the
doubling of shot noise, as found for normal/superconductor (NS) interfaces
theoretically in Refs. \cite{Bee,Muz,Naz} and experimentally confirmed in Refs.
\cite{Je1,Je2,Koz,Je3}.
Moreover, in Refs. \cite{An.Da,Ma,To.Ma.1,Gr.But,To.Ma.2} it has been shown
that multi-terminal
hybrid structures containing superconducting inclusions or leads can, in
some cases, exhibit
positive cross-correlations.
In particular, 
in Ref. \cite{An.Da} current correlations have been studied as a function
of the phase difference $\phi$ between the superconducting order
parameter of two superconducting pieces.
In the sub-gap regime,
positive cross-correlations have been found within a range of $\phi$
in the case of a floating superconducting electrochemical potential.
In Refs. \cite{Ma,To.Ma.1,To.Ma.2} a fork-like geometry with a
superconducting leg has been considered and the possibility of positive
cross-correlations has been found for bias voltages smaller than the superconducting energy gap.
In Ref. \cite{Gr.But} positive cross-correlations have been
predicted in a NS heterostructure with a weakly coupled tip.
It has been argued, however, that such
positive correlations vanish in the limit of large number of channels or
in the presence of disorder.

It is worthwhile mentioning that all
the above results are relative to a fully phase-coherent transport regime.
A doubling of shot noise, however, has been also found within a semi-classical
theory \cite{Nag1,Nag2}, therefore proving that phase-coherence is not required
for such an effect to occur.
More importantly, in hybrid systems in which a semi-classical treatment is valid,
cross-correlations have been demonstrated
to be always negative.

Recently, advances in nano-fabrication techniques of hybrid
normal-metal/ferromagnet structures has
led to the advent of spintronics \cite{Pri}.
In such a context
the spin degree of freedom of electrons is exploited (instead of their charge)
and spin-current correlations can be accessed.
In this paper we introduce this new ingredient and we
address the question whether ferromagnetism has an effect on the sign
of the cross-correlations.

As far as transport is concerned, a ferromagnet can be thought as a conductor
in which spin-up and spin-down electrons contribute by a different amount to the
total current (the spin degeneracy is broken).
For example, in the limit of a 100\% polarized ferromagnet (the so called
``half metal''), current is carried by a single spin species.
Due to the fact that in a s-wave superconductor spin-up particles are coupled
to spin-down holes, non-trivial effects are expected on the cross-correlations
between two ferromagnetic contacts having magnetizations in opposite directions.
We shall show that a positive value of cross-terminal correlations
in such hybrid systems can be induced
(as noticed in Ref. \cite{lesovik})
or, if already present, enhanced.
Remarkably, positive cross-correlations are predicted over a wide range of bias
voltages, even above the energy gap, for structures which exhibit negative
correlations when the ferromagnets are replaced with normal metals.
As expected, such an effect gets stronger as the ferromagnetic exchange field is
increased and reaches a maximum for exchange fields
corresponding to the limiting case of half-metallic leads.
The effect remains essentially unchanged even in the presence of disorder, 
due to impurities and lattice imperfections. 

\section{The model}

We consider a realistic hybrid 2D system
attached to three multi-channel terminals.
As sketched in Fig. \ref{dis1}, it consists of a
superconducting island (which defines the scattering region) connected to one normal metallic
lead (1) on the left-hand-side and two ferromagnetic leads (2 and 3) on 
the right-hand-side (similar structures have already been considered in
Refs. \cite{lesovik,Fei,Pino,Mel}).
The directions of magnetization of leads 2 and 3 are chosen to be anti-parallel
to each other.
The two ferromagnetic electrodes are kept at the same potential with respect to
the normal lead ($V_1-V_2=V_1-V_3\equiv V$).
For definiteness we assume the magnetization of lead 2 directed upward while
the magnetization of lead 3 directed downward.

Shot noise and cross-correlations are calculated within the multiple scattering
approach.
The scattering matrix $S$ completely characterizes a given structure (scatterer)
attached, through perfect conductors (leads), to $N$ reservoirs of particles.
$S$ is
defined via the asymptotic wave functions of the leads, known as scattering states.
When the scatterer contains superconducting regions, the scattering states
are solutions of the Bogoliubov-de Gennes equation \cite{deGenn}
\begin{equation}
\left( \begin{array}{cc} H_p & \Delta \\ \Delta^* & H_h \end{array} \right)
\left( \begin{array}{c} u(x) \\ v(x) \end{array} \right)=E
\left( \begin{array}{c} u(x) \\ v(x) \end{array} \right) ~.
\label{BdG}
\end{equation}
In Eq.(\ref{BdG}) $H_p$ is the Hamiltonian relative to the particle degrees of freedom,
$H_h$ is the Hamiltonian relative to hole degrees of freedom
and $\Delta$ is the superconducting order parameter, non-zero only in the
superconducting regions.
Let us consider, for example, a unit flux of particles originating
from the $j$-th reservoir.
The scattering state for the $j$-th lead (assuming for simplicity a single
open channel per lead) takes the form
\begin{equation}
\psi_j (x)=\left( \begin{array}{c} u(x) \\ v(x) \end{array} \right)=
\left( \begin{array}{c} \frac{e^{ik_jx}}{\sqrt{h\nu_j}}+ (r_0)_j
\frac{e^{-ik_jx}}{\sqrt{h\nu_j}} \\ (r_a)_j\frac{e^{iq_jx}}{\sqrt{h\omega_j}}
\end{array} \right) ~,
\label{sc_st1}
\end{equation}
whereas for the $i$-th lead, with $i\ne j$, it reads
\begin{equation}
\psi_i (x)=\left( \begin{array}{c} u(x) \\ v(x) \end{array} \right)=
\left( \begin{array}{c} (t_0)_{ij}\frac{e^{ik_ix}}{\sqrt{h\nu_i}} \\
(t_a)_{ij}\frac{e^{-iq_ix}}{\sqrt{h\omega_i}} 
\end{array} \right) ~.
\label{sc_s2t}
\end{equation}
Here $(r_0)_{j}$ and $(r_a)_{j}$ are, respectively, normal and Andreev
reflection amplitudes, while $(t_0)_{ij}$ and $(t_a)_{ij}$ are, respectively,
normal and Andreev transmission amplitudes.
In Eqs. (\ref{sc_st1}) and (\ref{sc_s2t}) $\nu_j$ ($\omega_j$) is the group
velocity in lead $j$ for particles (holes) and $k_j$ ($q_j$) is the wave-vector
for particles (holes).
$S$ is the matrix of scattering amplitudes of dimension $(2N\times 2N)$ defined
by
\begin{equation}
\ket{out}=S\ket{in}
\end{equation}
which relates the amplitudes of an incoming wave function ($\ket{in}$) to
the amplitudes of an outgoing wave function ($\ket{out}$).
We assume the convention that the upper $N$ elements of such vectors are relative
to particles and the lower ones to holes.

Following Refs. \cite{But.Bl,But.1,But.0} the noise power $\bar{s}_{il}$
is defined as the Fourier transform of the mean correlations of the current fluctuations
in leads $i$ and $l$:
\begin{equation}
\frac{1}{2} \langle \hat{\Delta I_i}(t) \hat{\Delta I_l}(t')+
\hat{\Delta I_l}(t') \hat{\Delta I_i}(t) \rangle ~,
\label{stilde}
\end{equation}
where $\hat{\Delta I_i}(t)=\hat{I_i}(t)-\langle \hat{I_i}\rangle$ is the current
fluctuation operator in lead $i$ at time $t$ and the angle brackets
$\langle \ldots \rangle$ denotes a statistical average.
By defining the time-dependent current operator as
\onecolm
\begin{equation}
\hat{I}_i (t)=\frac{2e}{h} \int_0^\infty dE ~dE' \sum_a e^{\frac{i}{\hbar}(E-E')t}
\sum_{j\beta b \atop j'\beta' b'} \hat{a}^\dagger_{j\beta b} (E)~
A^a_{j\beta b;j'\beta' b'}(i,E,E') ~\hat{a}_{j'\beta' b'} (E') ~,
\label{curr}
\end{equation}
one obtains the following expression for the zero-frequency
noise power~\cite{But.1}
\begin{equation}
\bar{s}_{il}(V,T)=\frac{2e^2}{h}\int_0^\infty dE \sum_{aa'}\sum_{j\beta 
b \atop j'\beta' b'}
A^a_{j\beta b;j'\beta' b'}(i,E,E) A^{a'}_{j'\beta' b';j\beta b}(l,E,E)
\left\{ f_{j\beta}(E) \left[ 1-f_{j'\beta'}(E) \right] +
f_{j'\beta'}(E) \left[ 1-f_{j\beta}(E) \right] \right\} ~.
\label{sil}
\end{equation}
In Eq. (\ref{curr}) $\hat{a}^\dagger_{j\beta b} (E)$ is the creation operator for
an incoming particle ($\beta=+1$) or hole ($\beta=-1$) with open channel index $b$
at energy $E$ originating from lead $j$.
The quantity $A^a_{j\beta b;j'\beta' b'}(i,E,E')$ is defined in terms of the scattering
matrix as
\begin{eqnarray}
 A^a_{j\beta b;j'\beta' b'}(i,E,E') &=&\delta_{ij}\delta_{ij'}\delta_{\beta e}
\delta_{\beta' e} \delta_{ba} \delta_{b'a} -
\delta_{ij}\delta_{ij'}\delta_{\beta h}
\delta_{\beta' h} \delta_{ba} \delta_{b'a} -
\nonumber \\
&&-S^*_{(i+a),(j\beta b)}(E)S_{(i+a),(j'\beta' b')}(E')+
S^*_{(i-a),(j\beta b)}(E)S_{(i-a),(j'\beta' b')}(E') ~,
\label{Agra}
\end{eqnarray}
\twocolm \noindent
\noindent
whereas $f_{j\beta}(E)$ is the Fermi distribution function for particles
($\beta=+1$) or holes ($\beta=-1$) at energy $E$ in reservoir $j$:
$f_{j\beta}(E)=[1+\exp(E-\beta (eV_j-\mu)/kT)]^{-1}$, $\mu$ being the
condensate electrochemical potential.
Note that $S_{(i\alpha a),(j\beta b)}$ is the scattering amplitude for a particle
($\beta =+1$) or a hole ($\beta =-1$)
originating in lead $j$ and relative to channel $b$ to be transferred into
lead $i$, channel $a$ as a particle ($\alpha =+1$) or a hole ($\alpha =-1$).

To evaluate the elements of the S-matrix we make use of an efficient
recursive Green's function technique \cite{Ste} which is based on the
tight-binding model.
We therefore describe the system under investigation
in terms of a 2-dimensional tight-binding Hamiltonian on a square lattice:
\begin{equation}
H=\sum_{\vec{l},\alpha} \ket{\vec{l},\alpha}\epsilon_{\vec{l},\alpha} \bra{\vec{l},
\alpha} + \sum_{\vec{l},\vec{k},\alpha,\beta} \ket{\vec{l},\alpha}
V_{\vec{l},\alpha,\vec{k},\beta} \bra{\vec{k},\beta} ~,
\label{tbH}
\end{equation}
where $\ket{\vec{l},\alpha}$ is either a particle ($\alpha =+1$) or a hole
($\alpha =-1$) degree of freedom of an atomic-like state (Wannier function)
centered at a site $\vec{l}$.
In the case of a perfect crystal without imperfections
we set the on-site energies as follow
\begin{equation}
\epsilon_{\vec{l},\alpha}=\alpha (\epsilon_0-h) ~,
\label{onsite}
\end{equation}
where $\epsilon_0$ is a real constant determining the position of the Fermi
energy and $h$, according to the Stoner model, is the ferromagnetic exchange field,
non-zero only in leads 2 and 3 (positive in lead 2 and negative in lead 3).
Note that the spin degrees of freedom are not coupled, since the magnetizations
in leads 2 and 3 are assumed to lie along the same direction.
For the hopping potentials we have
\begin{equation}
V_{\vec{l},\alpha,\vec{k},\beta}=\alpha V\delta_{\alpha,\beta} ~,
\end{equation}
with $\vec{l}$ and $\vec{k}$ first nearest neighbours, and
\begin{equation}
V_{\vec{l},+,\vec{l},-}=V_{\vec{l},-,\vec{l},+}^*=\Delta ~,
\end{equation}
since in local s-wave superconductors the particle-hole coupling is on-site.
Note that while $V$ is a real constant determining the band width,
the superconducting order parameter $\Delta$ is complex.
The presence of impurities and lattice imperfections is modeled as Anderson
disorder, whereby $\epsilon_0$ in (\ref{onsite}) is replaced by a uniformly
distributed random number in the range
$[\epsilon_0-W/2,\epsilon_0+W/2]$, $W$ being the disorder strength. 

The total Hamiltonian is block diagonal so that one can separately consider
the Hamiltonian relative to spin-up particles and spin-down holes from
the Hamiltonian relative to spin-down particles and spin-up holes.
Since they are decoupled, no correlations are present between spin-up and spin-down
particles (or holes).
From this follows that, as long as the system is symmetrical under the exchange of leads 2 and 3,
the results for spin-down polarized injected current are
equal to the results for spin-up polarized injected current once indices 2 and 3 are exchanged.
For simplicity, then, we assume the current injected from lead 1 to be 100\% spin-up
polarized
and for this reason we can drop the spin index in the definition of the
on-site energy.
In the recursive Green's function technique the scattering amplitudes are
calculated numerically from the total Green's function of the system, which in turn is
evaluated through the effective Hamiltonian of the scattering region
and the Green's functions of the isolated leads \cite{Ste}.

\section{Results}

In our calculations the scatterer (shaded region in  Fig. \ref{dis1}) is a
superconductor 15 sites long (in unit of lattice
spacings) and 15 sites wide.
The leads are 3 sites wide:
lead 1 is attached in the central position on the left-
hand-side, whereas leads 2 and 3 are placed symmetrically with respect to the center
of the right-hand-side at a distance of 5.
In arbitrary units, the order parameter is set to $\Delta=0.1$
uniformly in the superconducting region, therefore
neglecting the suppression of the superconducting order parameter in the
regions adjacent to the ferromagnets.
Such a suppression would lead to very small corrections, since it is
expected to occur
over very short lengths (units of nanometers for hard ferromagnets),
orders of magnitude smaller than the superconducting coherence length.
The temperature in
the reservoirs is set to $T=0.01$, while the Fermi energy is about $1.6$.
With these values we have three open channels in
the normal lead for particles and holes at the Fermi energy and
a superconducting
coherence length of the order of 10 lattice constants.
Note furthermore that the condensate electrochemical potential $\mu$ is
set to $\mu =0$.

As a reference situation we first consider the case where leads 2 and 3 are
normal-metallic ($h=0$).
In Fig. \ref{s.nor.nor} and Fig. \ref{s.nor.sup} we plot the noise power relative to the
different pairs of leads as a function of the bias voltage $V$.
When the superconductor is in the normal state (Fig. \ref{s.nor.nor}) we find, as
expected, that the shot noise in lead 1 is positive whereas the cross-correlations
are negative.
In the presence of superconductivity (Fig. \ref{s.nor.sup}), even though the magnitude of the
cross-correlations changes, the sign remains negative.
We now arrive at the main result of the paper, namely, that a positive sign can be
induced by replacing leads 2 and 3 with ferromagnets
characterized by magnetizations in opposite directions.
This is shown in Fig. \ref{s.half.sup} for the limiting case of half-metals,
where cross-correlations exhibit a positive sign.
While $\bar{s}_{12}$ changes sign as $V$ is increased, the effect
on $\bar{s}_{23}$
is more dramatic as it maintains a positive sign over the whole voltage range, with
a large magnitude.
It is remarkable that $\bar{s}_{23}$  remains positive over wide ranges of
voltages, even exceeding the superconducting energy gap.
Of course $\bar{s}_{12}$ and $\bar{s}_{13}$ are less affected
by ferromagnetism than $\bar{s}_{23}$, since lead 1 is normal-metallic.
In the following we will concentrate on the behaviour of $\bar{s}_{23}$.

In Fig. \ref{s23.vs.h} we plot $\bar{s}_{23}$ as a function of the magnitude
of the exchange field $h$ in the ferromagnetic leads 2 and 3 for four different values
of bias voltages.
First note that the four curves show the same behaviour, more pronounced in
magnitude for larger $V$.
For zero exchange field, {\em i.e.} with normal contacts, $\bar{s}_{23}$ takes
negative values despite the presence of superconductivity.
For finite values of $h$, $\bar{s}_{23}$ increases turning to positive
values, and thereafter
increasing further rapidly reaching a plateau.
The plateau corresponds to values of exchange field relative to
full polarization,
therefore it is an artifact of the small number of open channels.

As already noted in Refs. \cite{But.Bl,An.Da}, the reason for which
positive cross-correlations arise in the presence of superconductivity
can be understood by expressing the current operator of Eq. (\ref{curr}) as a
sum of particle and hole contributions:
\begin{equation}
\hat{I}_i (t)=\hat{I}^+_i (t)+\hat{I}^-_i (t) ~.
\label{curr.sum}
\end{equation}
By substituting Eq. (\ref{curr.sum}) in Eq. (\ref{stilde}), the noise power $\bar{s}_{il}$
[Eq. (\ref{sil})] can be written as the following sum:
\begin{equation}
\bar{s}_{il}=\bar{s}_{il}^{++}+\bar{s}_{il}^{--}+\bar{s}_{il}^{+-}+
\bar{s}_{il}^{-+}
\label{sums}
\end{equation}
where $\bar{s}_{il}^{++}$ is the particle-particle (pp) cross-correlation,
$\bar{s}_{il}^{+-}$ is the particle-hole (ph) cross-correlation and so on.
Quite generally it was shown in Ref. \cite{An.Da} that while 
$\bar{s}_{il}^{++}$ and
$\bar{s}_{il}^{--}$ are negative,
$\bar{s}_{il}^{+-}$ and $\bar{s}_{il}^{-+}$ take a positive sign.
Intuitively, one can understand that pp (and hh) correlations are negative because 
of the anti-bunching behaviour due to the Fermionic nature of electrons 
\cite{But.Bl,But.1}.
For the same reason, ph and hp correlations present a positive sign
simply because particle and hole have opposite charge
(note that when the superconductor is in the normal state ph and hp
correlations are zero).
The sign of the overall cross-correlation is therefore determined
by the competition between the positive (ph and hp) and the negative (pp 
and hh) contributions.
The important observation is that pp and hh cross-correlations can
be drastically suppressed in the presence of spin-polarized transport, for
example by allowing only spin-up polarized current in one lead and only
spin-down polarized current in the other.
As a result an overall positive sign will be induced.

In the system under investigation the different contributions to $\bar{s}_{23}$,
in the zero-temperature, small bias voltage limit and with
$\mu$ set to zero, take the simple form
\begin{equation}
\bar{s}_{23}^{\alpha\gamma}=-(\alpha\gamma) \frac{4e^3V}{h} \left[  
T_{21}^{\alpha +}
T_{31}^{\gamma +} +T_{21}^{\alpha -} T_{31}^{\gamma -} \right] ~,
\label{svtzero}
\end{equation}
where $T_{i1}^{\alpha\beta}$ is the transmission probability, evaluated at
$E=0$, for a particle ($\beta =+1$) or a hole ($\beta =-1$) injected from
lead 1 to be transmitted into lead $i$ as a particle 
($\alpha =+1$) or a hole ($\alpha =-1$).
In the limiting situation where leads 2 and 3 are half-metallic with 
magnetizations aligned anti-parallel to each other (for example upward for 
2 and downward for 3), we have that only positive correlations 
are non-zero.
Particle-particle and hole-hole correlations, in fact, vanish
($\bar{s}_{23}^{++}=\bar{s}_{23}^{--}=0$) because transmission of
spin-up particles is forbidden in lead 3 and transmission of
spin-down particles is forbidden in lead 2:
$T_{31}^{+\alpha}=T_{21}^{-\alpha}=0$.
For weaker ferromagnets $\bar{s}_{23}^{\alpha\alpha}$ are finite, but 
nevertheless suppressed with respect to ph correlations so 
that $\bar{s}_{23}$ remains positive.
It is worthwhile noting that it is possible to switch the sign of
$\bar{s}_{23}$ from positive
to negative by changing the relative alignment of the magnetizations in the two
ferromagnetic leads from anti-parallel to parallel.
In the latter case, in fact, $T_{31}^{-\alpha}$ and $T_{21}^{-\alpha}$ in Eq.
(\ref{svtzero}) are both equal to zero in the limit of half-metallic leads,
so that only $\bar{s}_{23}^{++}$ is left finite,
analogously to what happens
when the superconductor is turned normal.
For completeness, note that by summing over the indices $\alpha$ and 
$\gamma$ in Eq. (\ref{svtzero}) and employing the particle-hole
symmetry one obtains the simple expression
\begin{eqnarray}
\bar{s}_{23}=-\frac{8e^3V}{h}\left( |(t_0)_{31}|^2 
- |(t_a)_{31}|^2 \right) \times \nonumber \\
\times \left( |(t_0)_{21}|^2 - 
|(t_a)_{21}|^2 \right) ~.
\label{s23}
\end{eqnarray}

We can also understand the reason why positive correlations in 
Fig. \ref{s.half.sup} survive over such a wide voltage range.
While pp correlations are zero, ph correlations tend 
to vanish only for quasi-particles energy well above the superconducting gap 
(about one order of magnitude),
{\em i.e.} when Andreev processes are negligible.
Furthermore, the effect on the sign of $\bar{s}_{12}$ and $\bar{s}_{13}$ is
less pronounced than on $\bar{s}_{23}$, because in the former cases pp or hh
correlations can still be large.
It is worthwhile mentioning that the presence of barriers at the
interfaces between electrodes and the
superconductor is expected to affect the cross-correlations
(analogously to the findings of Ref. \cite{To.Ma.1}).
Eq. (\ref{s23}) shows that the sign of $\bar{s}_{23}$ can change with
the barrier strength,
since such a sign is determined by the interplay between
normal and Andreev transmissions between leads 1 and 2, and 1 and 3.
In the case of half-metallic contacts,
it is clear that cross-correlations $\bar{s}_{23}$ will still be positive
regardless of the value of the barrier strength, because negative
contributions (pp and hh correlations) are always zero.
However, we expect that the magnitude of $\bar{s}_{23}$ might
depend on the barrier strength non-monotonically.
In fact, while normal transmission coefficients decrease with
increasing barrier strength, Andreev transmissions do not monotonically
depend on the barrier strength, but instead first
increase and then decrease.

To conclude, we have also checked that the presence of weak disorder within the
superconductor, due to
impurities or lattice imperfections, does not significantly change the results.
Using the Anderson model described above, we have considered the same system studied
in the clean limit for different values of the disorder strength $W$.
Even without performing a systematic study of disordered systems, we have
found that $\bar{s}_{23}$ always remains positive over the whole
voltage range, though the shape of the curve changes.
On the contrary, for some disorder realizations we have found that $\bar{s}_{12}$ and
$\bar{s}_{13}$ do not change sign remaining negative over the voltage range considered.
For the future it would be of interest a systematical study of disordered
systems in the different transport regimes.
We argue that even in the diffusive regime and with a large number of open channels
in the contacts such positive sign survives.

\section{Summary}
The study of electronic current-current correlations in mesoscopic structures
has been receiving interest since the last decade.
On the one hand, it has been found that the correlations between
different leads are always negative.
On the other, in hybrid NS structures it has been proved that such correlations
can, in some cases, be positive.
Here we have considered the sign of cross-correlations in a hybrid NS
three-terminal structure in which ferromagnetic leads are employed.
We have found that a positive sign of the cross-correlations between two ferromagnetic
leads is induced when their magnetizations are aligned
anti-parallel to each other.
It is remarkable that such a positive sign persists over a wide range of voltages
exceeding the value corresponding to the superconducting energy gap.
Furthermore positive cross-correlations have also been found between normal metallic and
ferromagnetic leads.
The origin of this effect has been attributed to the suppression of particle-particle
and hole-hole cross-correlations (which contribute with a negative sign) in favour of
particle-hole correlations (which contribute with a positive sign), due to the
strong spin-asymmetry introduced by ferromagnetism.
We have moreover realized that it is possible to switch the sign of $\bar{s}_{23}$ from
positive to negative by changing the the relative alignment of the magnetizations
in the two ferromagnetic leads from anti-parallel to parallel.
Such effects are robust against the presence of weak disorder
within the superconducting region.
To conclude, we would like to enphasize that our approach can be applied to disordered
diffusive samples as well as clean and ballistic ones and to any geometry.
Furthermore, it can be easily extended to finite frequencies.


\acknowledgments 
The authors would like to thank C. J. Lambert and G. Falci for helpful 
discussions. This work has been supported by the EU (IST-FET-SQUBIT) and  
by INFM-PRA-SSQI.


\begin{figure}
\begin{center}
\epsfig{figure=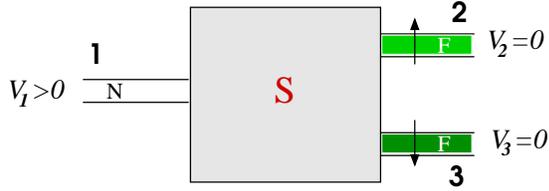,width=0.4 \textwidth}
\end{center}
\caption{The system consists of a superconductor (S) connected to 3
contacts. Terminals 2 and 3 are made of ferromagnets, with
anti-parallel magnetizations, while terminal 1 is normal-metallic.
Lead 1 is kept at positive potential $V_1$ with respect to leads 2 and 3.
The relevant geometric lengths are of the same order of the superconducting
coherence length.}
\label{dis1}
\end{figure}

\begin{figure}
\begin{center}
\epsfig{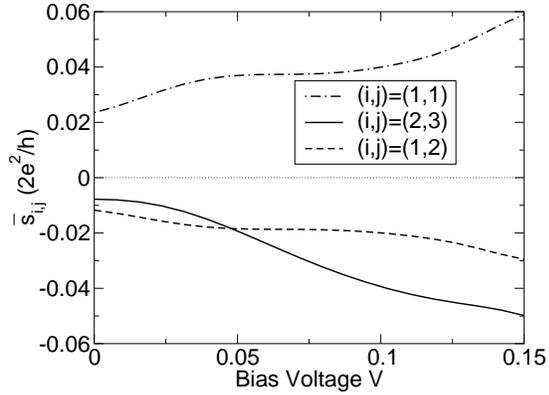}
\end{center}
\caption{Noise power as a function of bias voltage $V$ with S in the normal state
and $h=0$ in leads 2 and 3. Due to the geometrical symmetry $\bar{s}_{13}$ is
exactly equal to $\bar{s}_{12}$.}
\label{s.nor.nor}
\end{figure}

\begin{figure}
\begin{center}
\epsfig{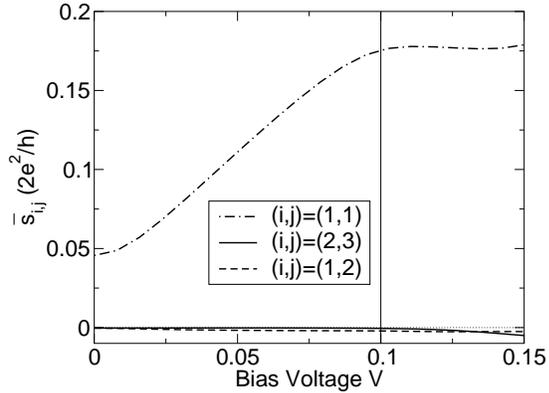}
\end{center}
\caption{Noise power as a function of bias voltage $V$ with S in the superconducting state
and $h=0$ in leads 2 and 3. The vertical line denotes the voltage
corresponding to the superconducting energy gap.
Due to the geometrical symmetry $\bar{s}_{13}$ is
exactly equal to $\bar{s}_{12}$.}
\label{s.nor.sup}
\end{figure}

\begin{figure}
\begin{center}
\epsfig{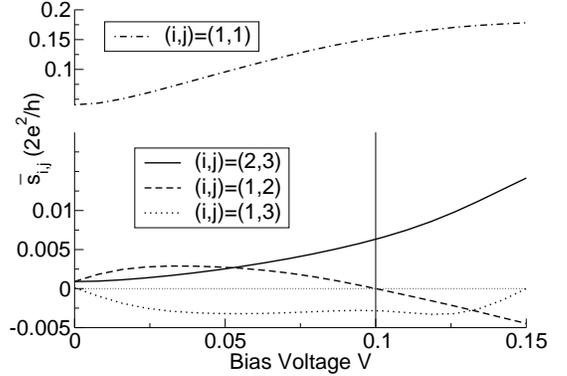}
\end{center}
\caption{Noise power as a function of bias voltage $V$ with S in the superconducting state
and half-metallic leads 2 and 3. Note that cross-correlations are plotted on a
different scale with respect to $\bar{s}_{11}$.
The vertical line denotes the voltage
corresponding to the superconducting energy gap.}
\label{s.half.sup}
\end{figure}

\begin{figure}
\begin{center}
\epsfig{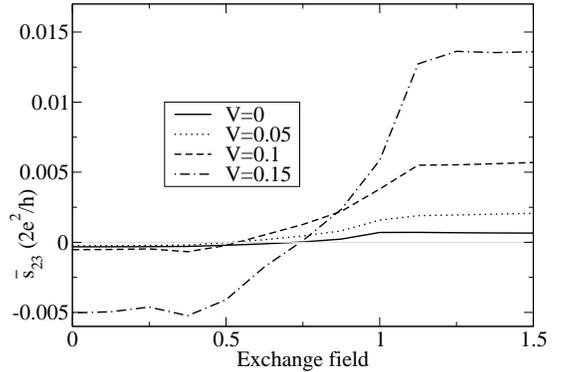}
\end{center}
\caption{Cross-correlation $\bar{s}_{23}$ as a function of the magnitude of the exchange
field $h$ in leads 2 and 3 for four different bias voltages.}
\label{s23.vs.h}
\end{figure}

\end{multicols}

\end{document}